\documentclass[11pt,a4paper, twocolumn]{aa}
\usepackage[utf8]{inputenc}
\usepackage[T1]{fontenc}
\usepackage{lmodern}
\usepackage{microtype}
\usepackage{setspace}
\usepackage{graphicx}
\usepackage{booktabs}
\usepackage{hyperref}
\hypersetup{colorlinks=true,linkcolor=blue,citecolor=blue,urlcolor=blue}
\usepackage{titling}
\usepackage{fancyhdr}
\usepackage{amsmath,amssymb}
\usepackage{multicol}

\begin{document}

\begin{titlepage}
  \centering
  \vspace*{1cm}
  {\Huge\bfseries Expanding Horizons \\[6pt] \Large Transforming Astronomy in the 2040s \par}
  \vspace{1.5cm}

  {\LARGE \textbf{White Dwarfs in Wide Binary Systems as Reliable Age Calibrators}\par}
  \vspace{1cm}

  \begin{tabular}{p{4.5cm}p{11cm}}
    \textbf{Submitting Author:} & Alberto Rebassa-Mansergas$^{1,2}$ \\
    & Email: alberto.rebassa@upc.edu\\
    \\
    \textbf{Contributing authors:} & Roberto Raddi$^{1}$, Anna F. Pala$^{3}$, Alejandro Santos-García$^{1}$, Santiago Torres$^{1,2}$, Leandro Althaus$^{4,1}$, Diogo Belloni$^{5}$, Maria Camisassa$^{1}$, Tim Cunningham$^{6}$, Camila Damia Rinc\'on$^1$, Aina Ferrer i Burjachs$^{1}$, Enrique Garc\'ia-Zamora$^1$, JJ Hermes$^{7}$, Adam Moss$^{8}$, Steven Parsons$^{9}$, Odette Toloza$^{10}$  \\\\

\multicolumn{2}{l}{\small $^{1}$Departament de Física, Universitat Politècnica de Catalunya, c/Esteve Terrades 5, 08860, Castelldefels, Spain} \\
\multicolumn{2}{l}{\small $^{2}$Institut d'Estudis Espacials de Catalunya (IEEC), C/Esteve Terradas, 1, Edifici RDIT, 08860, Castelldefels, Spain} \\
\multicolumn{2}{l}{\small $^{3}$European Southern Observatory, Karl Schwarzschild Straße 2, D-Garching
85748, Germany}\\
\multicolumn{2}{l}{\small $^{4}$Facultad de Ciencias Astronómicas y Geofísicas, Universidad Nacional de La Plata, Paseo del Bosque s/n, 1900 La Plata, Argentina}\\
\multicolumn{2}{l}{\small $^{5}$São Paulo State University (UNESP), School of Engineering and Sciences, Guaratinguetá, Brazil}\\
\multicolumn{2}{l}{\small $^{6}$Center for Astrophysics, Harvard \& Smithsonian, 60 Garden St., Cambridge, MA 02138, USA}\\
\multicolumn{2}{l}{\small $^{7}$Department of Astronomy, Boston University, 725 Commonwealth Avenue,
Boston, MA, 02215, USA}\\
\multicolumn{2}{l}{\small $^{8}$Department of Astronomy, University of Florida, Bryant Space Science Center, Stadium Road, Gainesville, FL 32611, USA}\\
\multicolumn{2}{l}{\small $^{9}$Department of Physics and Astronomy, University of Sheffield, Sheffield S3
7RH, UK}\\
\multicolumn{2}{l}{\small $^{10}$Departamento de Física, Universidad Técnica Federico Santa María, Avenida España 1680, Valparaíso, Chile}\\  
  \end{tabular}

  \vspace{1cm}

  \vspace{0.5em}
  \begin{minipage}{0.9\textwidth}
    \small

   \textbf{Abstract:} 

Deriving precise stellar ages is a challenging task. Consequently, age-dependent relations—such as the age–metallicity and age–velocity dispersion relations of the Milky Way, or the age–rotation–activity relation of low-mass stars—are subject to potentially large uncertainties, despite the well-defined trends observed at the population level.

White dwarfs, the most common stellar remnants, follow a relatively simple and well-understood cooling process. When found in wide binary systems with main-sequence companions, they can therefore provide the much-needed precise age estimates. The total age of such systems depends not only on the white dwarf cooling time but also on the lifetime of the main-sequence progenitor. Estimating this lifetime requires knowledge of the progenitor mass, which is typically inferred by adopting an initial-to-final mass relation. However, the observational constraints on this relation are still poorly defined, introducing a source of uncertainty in white dwarf age determinations.

To mitigate this issue, we focus on a large sample of massive white dwarfs ($\gtrsim 0.7\,M_{\odot}$), for which the main-sequence progenitor lifetime is negligible. These white dwarfs are intrinsically faint and therefore require specialized facilities for adequate follow-up observations. In this white paper, we outline the instrumentation requirements needed to observe the forthcoming population of massive white dwarfs in our Galaxy.

  \end{minipage}

\end{titlepage}


\section{Introduction and Background}
\label{sec:intro}

Measuring stellar ages is of paramount importance to constrain a wide variety of open questions in modern astronomy such as: 
\vspace{0.1cm}

    \underline{The age–metallicity relation}. It describes how the chemical enrichment of the Galaxy varies with stellar age. Early work established that older stars tend to be more metal-poor \citep{Twarog1980, Edvardsson1993}, consistent with gradual enrichment of the interstellar medium. However, observations show substantial intrinsic scatter in the solar neighborhood, a feature confirmed by modern high-precision photometric and spectroscopic surveys \citep{Casagrande2011, Bergemann2014}. This scatter is now understood to potentially reflect processes such as radial migration and spatially varying star-formation histories, as claimed by recent analyses of APOGEE and $Gaia$ samples \citep{Feuillet2018, Sahlholdt2022}.

    \underline{The age–velocity dispersion relation}. It captures the empirical increase in stellar velocity dispersion with age in the Galactic disk. Early studies showed that older stars have systematically larger dispersions in all velocity components \citep{Wielen1977}, a trend confirmed in larger kinematic samples such as the Geneva–Copenhagen Survey \citep{Nordstrom2004}. This behavior is interpreted as the result of progressive dynamical heating by molecular clouds, spiral structure, and/or other perturbations \citep{Aumer2009}. With the advent of $Gaia$, the age-velocity dispersion relation has been mapped with much greater precision, revealing changes in the heating rate and signatures of past dynamical events \citep{Ting2019, Sharma2021}.

    \underline{The age–activity–rotation relation}. It describes the coupled evolution of rotation, magnetic activity, and age in low-mass stars. The foundational observation that rotation and chromospheric activity decline roughly as $t^{-1/2}$ established the basis for rotational aging \citep{Skumanich1972}. This framework was later formalized into modern gyrochronology relations linking rotation period, activity, and age \citep{Barnes2007, Mamajek2008}. 
    More recent work has revealed complexities in this evolution, including weakened magnetic braking and deviations from classical gyrochronology at older ages \citep{vanSaders2016, Curtis2020}, as well as improved calibrations enabled by large rotation samples \citep{Lu2024}.
\vspace{0.1cm}

Despite major advances enabled by large photometric, spectroscopic, and astrometric surveys, obtaining accurate and precise ages for individual field stars remains intrinsically challenging \citep{Soderblom2010}. Systematic model assumptions, degeneracies in stellar parameters, and the intrinsic scatter in empirical relations can all lead to substantial age errors of several gigayears for single stars, even when population-level trends seem to be well defined \citep{Nordstrom2004, Jorgensen2005, Ting2019, Curtis2020}.

White dwarfs are the most common stellar remnants and, once they are formed, they follow a relatively well understood cooling process \citep{Althaus2010, camisassa2016, camisassa2019}. Therefore, they hold the potential to provide the much-needed accurate ages to the astronomical community, with precisions ranging from $\simeq5-25\%$, depending on the white dwarf mass \citep{2019Fouesneau, Moss2022, Heintz2022}. Of special interest are those white dwarfs that are members of binary star systems with main-sequence companions. If the binary components are separated enough ($\ga$10 AU; \citealt{Farihi2010}) this ensures the systems very likely evolved avoiding mass transfer episodes. Thus, one can safely assume that the measured white dwarf ages are the same as those of the main-sequence companions, since the two stars are coeval. These white dwarf plus main-sequence (WD+MS) binaries can be used as ``tools'' to analyze any age-related open issue. 

Thanks to the astrometry and photometry data provided by the $Gaia$ satellite \citep{Gaia2023}, a relatively large number of WD+MS binaries has been identified in common proper motion pair systems, that is, widely-separated binaries \citep[e.g.][]{ElBadry2021}. This allowed dedicated observational studies to constrain both the chemical and dynamical evolution of the Milky Way through the age-metallicity and age-velocity dispersion relations, as well as the relation between age, activity and rotation of low-mass stars \citep[e.g.][]{Rebassa2021, Rebassa2023, Raddi2022, Chitti2024}. These pilot studies agree well with the results obtained from single isolated stars, strengthening our understanding of such important relations. From 2026 to the beginning of the 2030s, the White Dwarf Binary Survey of 4MOST (the 4-metre Multi-object Spectroscopic Telescope; \citealt{dejong2019}) will provide white dwarf ages as well as main-sequence star metallicities, radial velocities, rotational velocities and activity indexes for $\simeq$3000 WD+MS binaries \citep{Toloza2023}. This will increase the current number of well-characterized WD+MS systems with age determinations by one order of magnitude.

It is important to remark that, despite the important contribution of WD+MS binaries to the aforementioned studies, white dwarf total ages are the sum of their cooling times plus their main-sequence progenitor lifetimes. This introduces uncertainties in the derived white dwarf ages because of the following reasons:
\vspace{0.1cm}

    On the one hand, there is still no consensus on the observational properties of the initial-to-final mass relation \citep[IFMR;][]{Cummings2018, Ironi2025}. 
    The IFMR allows the masses of main-sequence stars to be inferred from measurements of present-day white dwarf masses. These estimates can then be used, together with evolutionary sequences and a known metallicity, to determine the corresponding main-sequence lifetimes. Consequently, the inferred lifetimes of main-sequence progenitors depend sensitively on the adopted IFMR, which can vary substantially \citep{Rebassa2016}. Moreover, the lower the white dwarf mass, the longer its progenitor remained on the main sequence. Because the main-sequence lifetime derived from the IFMR is highly sensitive to the progenitor mass, even small errors in the masses of low-mass white dwarfs lead to significantly different main-sequence lifetimes and, therefore, to substantially larger uncertainties in the total ages.

    On the other hand, the metallicities of white dwarfs are unknown due to the rapid diffusion of elements heavier than H and He into their deep interiors. As a result, the metallicities of their progenitors are unconstrained, further increasing the uncertainty when adopting an IFMR. Fortunately, in WD+MS binaries, the metallicity of the main-sequence companion can be used as a proxy for that of the white dwarf progenitor.
\vspace{0.1cm}

While the wide binaries with measurable ages for the non-degenerate companions can help to constrain the IFMR  \citep{Catalan2008, Barrientos2021}, we require a large sample of WD+MS binaries containing relatively massive white dwarfs ($\gtrsim 0.7,M_{\odot}$). Such white dwarfs have very short main-sequence progenitor lifetimes \citep{camisassa2016, camisassa2019}, allowing the total system age to be approximated safely by the white dwarf’s cooling age, which is considerably more reliable and independent of metallicity. However, massive white dwarfs are not only rare but also intrinsically faint due to their small radii and their faster cooling. Consequently, they constitute only a rather small fraction of the total white dwarf and WD+MS binary populations in volume-limited samples \citep{McCleery2020, Jimenez2023, Kilic2025} and, at the same time, they are harder to detect in magnitude-limited samples \citep{Rebassa2015, Torres2023}. Even within the 4MOST survey only a small fraction ($\simeq10$\%) of WD+MS binaries are expected to contain massive white dwarfs.


In the coming years, the Legacy Survey for Space and
Time (LSST) at the Vera C. Rubin Observatory \citep{Ivezic2019} will enable the identification of this elusive population (see the next section). However, no current or planned facility is capable of efficiently characterizing this sample. The purpose of this white paper is to outline the requirements for a facility that can meet this need. In Section\,\ref{s-sample} we provide the expected number of WD+MS binaries containing massive white dwarfs that LSST will be able to identify and in Section\,\ref{s-require} we discuss the technology and data handling requirements.

\section{The sample of WD+MS binaries containing massive white dwarfs}
\label{s-sample}

We used the Monte Carlo simulator MRBIN described in \citet{Santos2025} to quantify the total number of WD+MS binaries in the Galaxy up to 3000\,pc, and to evaluate how many of these have massive white dwarfs and will be accessible by LSST. The code not only is capable of reproducing realistic single main-sequence and single white dwarf Galactic populations \citep[e.g.][]{Torres2001, Torres2019}, but also allows to reproduce main-sequence binary stars and their subsequent evolution \citep{Torres2022}. 

The binary stellar evolution is modeled following the binary stellar evolution (BSE) code developed by \citet{Hurley2002}, with particular updates on white dwarf binaries \citep{Camacho2014, Cojocaru2017, Canals2018}. The modeling of the entire single and binary population requires a series of input parameters such as binary fraction, initial mass function, secondary mass function, star formation rate, etc. adopted from \citet{Torres2022}. Moreover, the simulator randomly assigns each single and binary star to a Galactic component 
following \citet{Torres2019}. Objects from the different Galactic components are modeled following different criteria regarding their ages, metallicities, spatial distributions and kinematics. 

Wide binary systems evolve in the same way as single stars, avoiding mass transfer episodes. However, for the evolution of binary systems with shorter separations, our code, based on the BSE code, uses a wide set of parameters to model the mass transfer, common envelope evolution and angular momentum losses. 

Once the systems (both single and binary stars) are evolved to the present time, the MRBIN code allows the derivation of magnitudes in a given photometric system ($Gaia$ and LSST in this case). For those systems that become white dwarfs, their cooling time is accurately determined using the most up-to-date evolutionary cooling sequences provided by the La Plata group \citep{camisassa2016, camisassa2019, althaus2025}, which take into account the different core and chemical compositions over the full mass range, as well as key physical processes such as crystallization, phase separation, updated opacities, and neutrino emission.

Each object was assigned an extinction based on their positions and distances, interpolated in the 3D maps of \citet{lallement2022}. Finally, we distinguished between
resolved and unresolved synthetic binaries as follows: we considered a binary to be unresolved if its angular separation was smaller than 1", which is slightly larger than the average seeing where the LSST will be operating. 

\begin{figure*}
    \centering
    \includegraphics[angle=-90,width=0.4\linewidth]{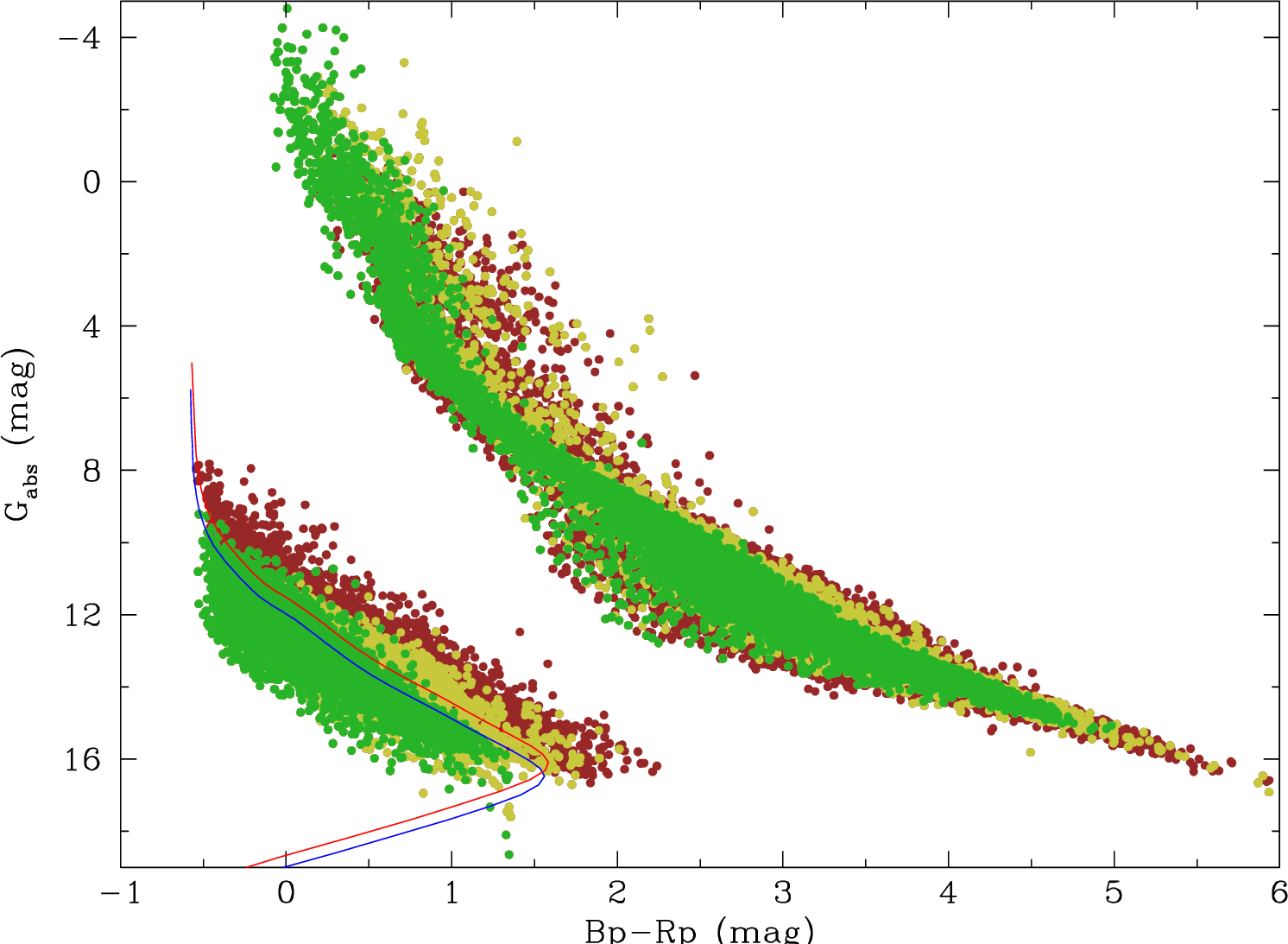}
    \includegraphics[angle=-90,width=0.4\linewidth]{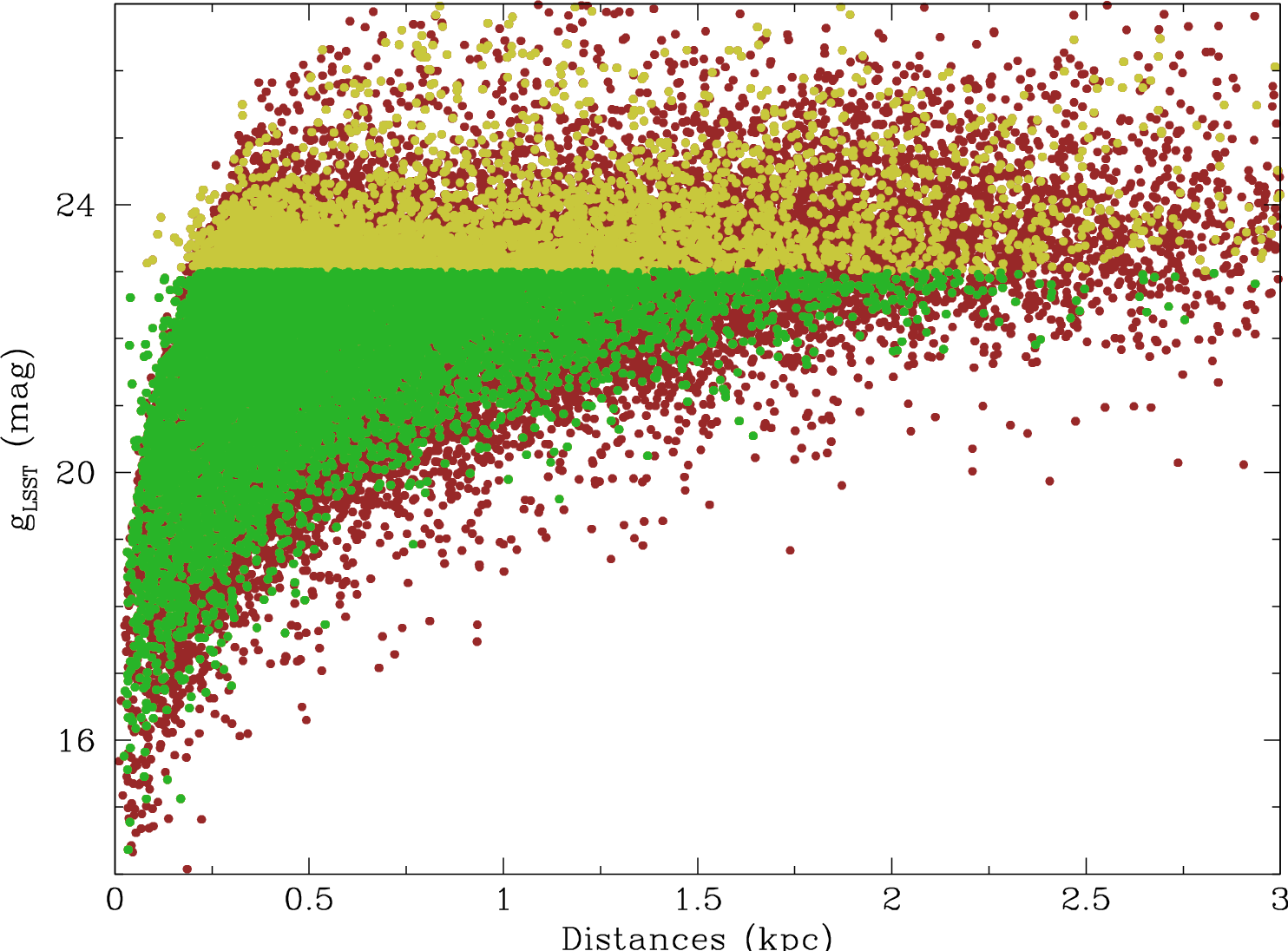}
    \caption{Left: $Gaia$ DR3 HR diagram for the synthetic WD+MS binaries within 3 kpc. Dark red shows all systems accessible by LSST ($\simeq$40\,500), dark yellow those containing massive white dwarfs ($M\geq0.7$\,M$_{\odot}$; $\simeq13\,000$ objects) and dark green those with g$_\mathrm{LSST}\leq23$ ($\simeq8\,600$ objects). The solid red and blue lines indicate the cooling sequences of a 0.5 and a 0.7 M$_{\odot}$ white dwarf, respectively. The synthetic massive white dwarfs (yellow, green) are spread above and below these limits due to extinction and parallax uncertainties considered by our code. Right: their $g_\mathrm{LSST}$ magnitudes as a function of distance.}
    \label{f-surveys}
\end{figure*}

Figure\,\ref{f-surveys} (left panel) illustrates the $Gaia$ Hertzsprung-Russell (HR) diagram for our simulated WD+MS binaries. In dark red all the systems ($\simeq$40\,500) accessible by LSST (declination less than 5 degrees), in dark yellow those containing white dwarfs with $M\geq0.7$\,M$_{\odot}$ ($\simeq13\,000$ objects) and in dark green those with g$_\mathrm{LSST}\leq23$ ($\simeq8\,600$ objects), which we interpret as the conservative magnitude limit of LSST. In the right panel of the same figure, we show the g$_\mathrm{LSST}$ magnitude as a function of distance. It can clearly be seen that most white dwarfs above 2.5\,kpc are fainter than 23 mag and that the vast majority of massive white dwarfs ($M\geq0.7$\,M$_{\odot}$) are located below 1-1.5\,kpc.

\begin{figure*}
    \centering
    \includegraphics[angle=-90,width=0.4\linewidth]{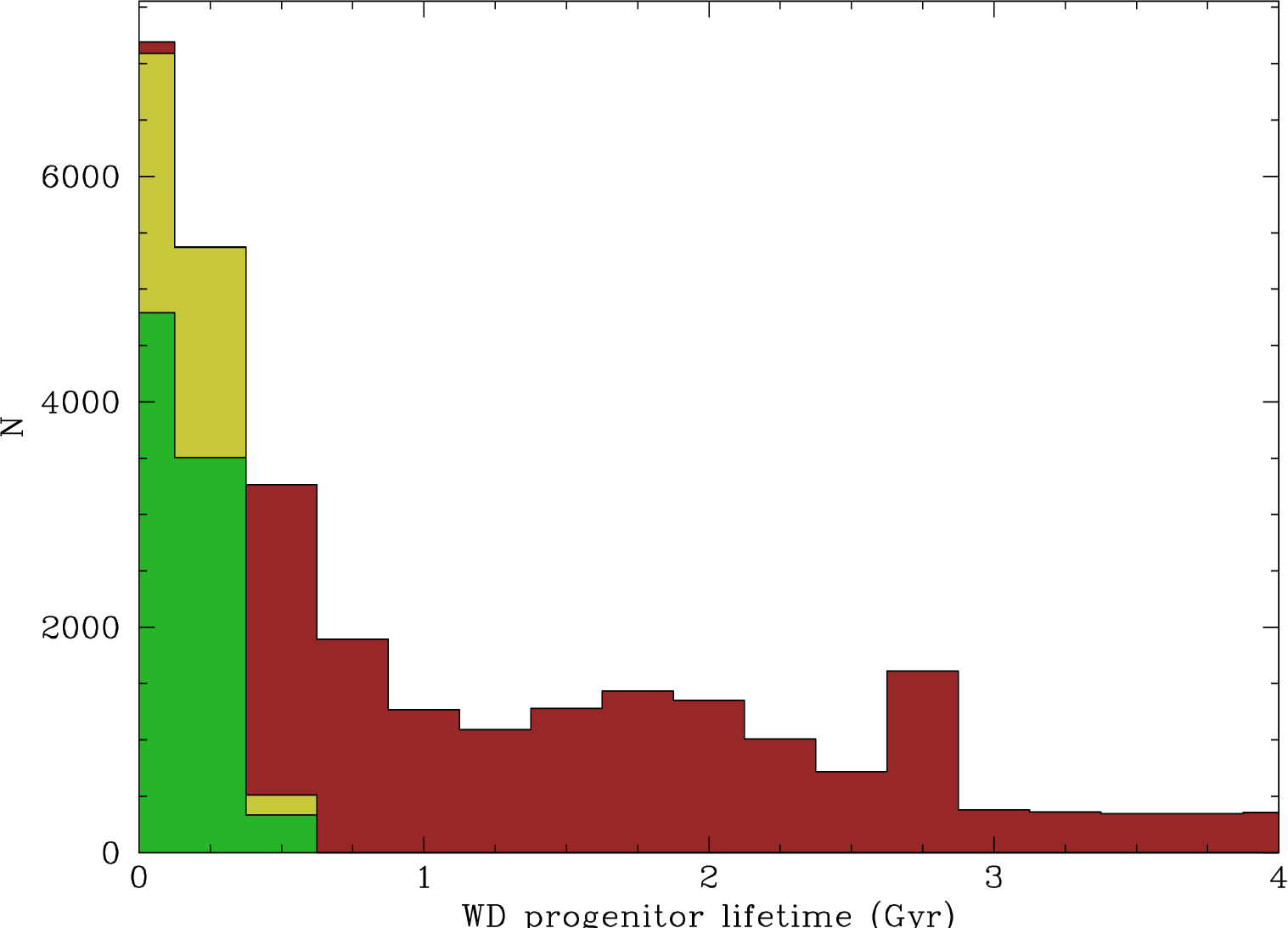}
    \includegraphics[angle=-90,width=0.4\linewidth]{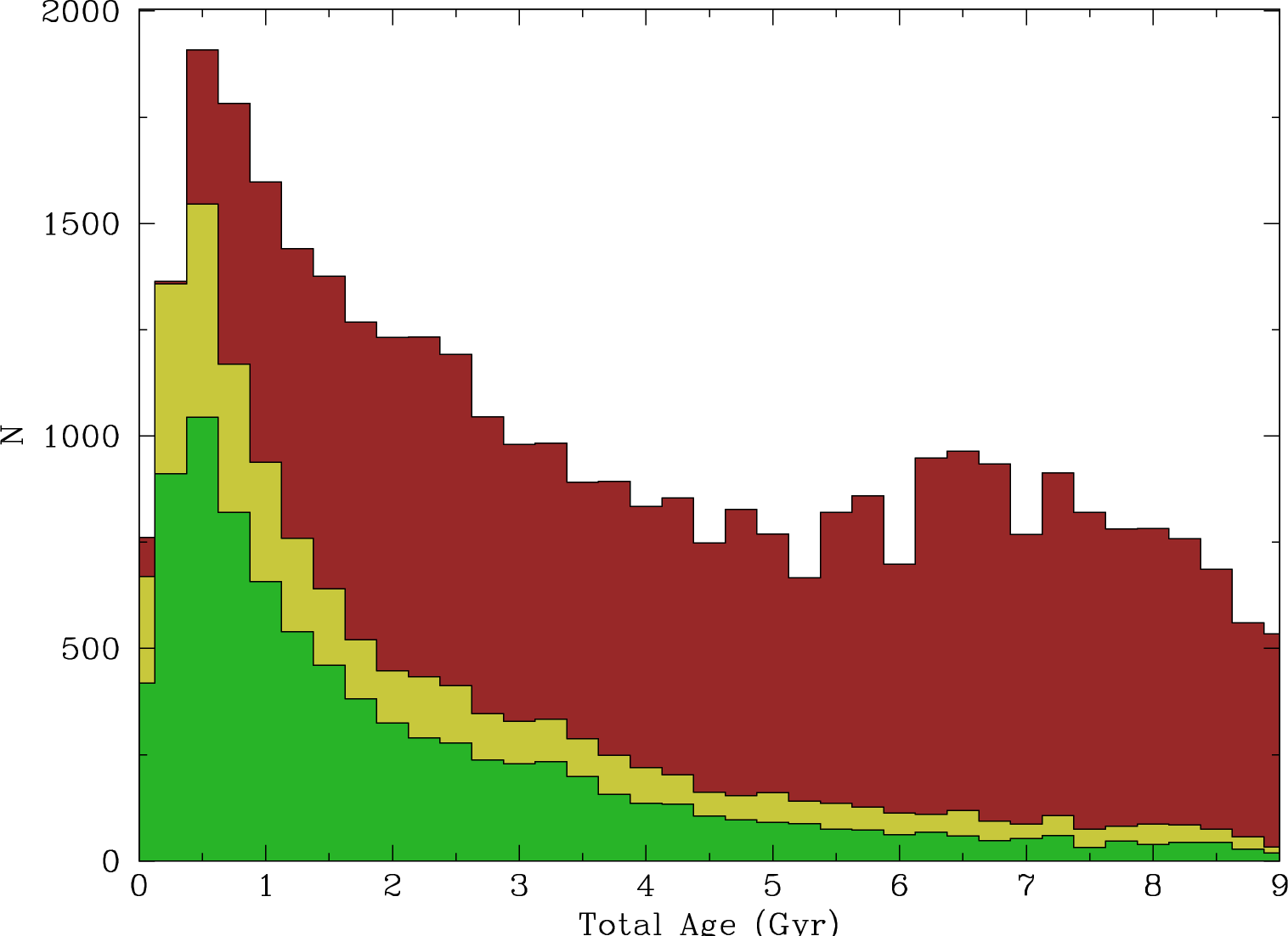}
    \caption{White dwarf progenitor (left) and total (right) ages for the synthetic WD+MS binaries. Colors are as in Figure\,\ref{f-surveys}.}
    \label{f-ages}
\end{figure*}

In Figure\,\ref{f-ages} we display the white dwarf progenitor lifetime (left) and total age (right) distributions for the three mentioned samples following the same color criteria. As it can be seen, all white dwarfs with masses larger than $0.7$\,M$_{\odot}$ have very short main sequence lifetimes ($\la 0.5$ Gyr) and sample all possible total ages (0-9 Gyr; although the majority are concentrated between 0-2 Gyr). Thus, this sample of $\simeq8\,600$ WD+MS binaries will enable unprecedented constraints on the evolution of low-mass stars and the assembly history of the Milky Way, while providing a foundation for a wide range of Galactic archaeology studies.


\bibliographystyle{aa}


\newpage
\section{Technology and Data Handling Requirements}
\label{s-require}

In order to fully characterize the population of massive white dwarfs in WD+MS binaries, a large-aperture telescope (or equivalent facility) capable of delivering a signal-to-noise ratio of $\ga 20$ in relatively short exposures is essential. Such sensitivity is required to obtain high-quality spectra of faint, high-mass white dwarfs, which are critical for constraining their cooling ages. In the coming years, the Extremely Large Telescope (ELT; 39 m), the Thirty Meter Telescope (TMT; 30 m), and the Giant Magellan Telescope (GMT; 21.4 m) will provide the necessary photon-collecting power. However, these facilities are not survey telescopes and are not optimized for projects that require spectra of thousands of stars. Their designs emphasize extremely high spatial resolution, deep targeted observations, and precision spectroscopy, rather than large-scale multiplexed surveys.

Among smaller-aperture telescopes that remain sufficiently large for high-mass white dwarf observations, the Maunakea Spectroscopic Explorer (MSE; 11.25 m) and the Subaru Prime Focus Spectrograph (PFS; 8.2 m) are particularly notable. MSE is currently on hold, with no active schedule for construction or operations, while PFS began operations in February 2025. However, PFS is not a dedicated white dwarf survey and Europe is not a formal partner.

The requested new facility should combine large aperture ($\ga 10$ m; or equivalent facility) with multi-object spectroscopic capabilities, enabling the efficient characterization of statistically significant samples of white dwarfs and their main-sequence companions. For the white dwarfs themselves, a minimum resolving power of $R\simeq2\,000$ is necessary to fit model atmospheres accurately, measure effective temperatures and surface gravities, hence constrain masses and cooling (i.e. total) ages. To that end full optical plus near-infrared coverage would be required ($\simeq3\,600-10\,000$\,\AA). Since the main-sequence companions must also be characterized — to measure metallicities, radial and rotational velocities, and trace Galactic kinematics — a higher-resolution mode ($R \ga 15\,000$) is required. In this case, the following three spectral regions would be sufficient (following the 4MOST structure): 3926–4355\,\AA\, (blue), 5160–5730\,\AA\, (green) and 6100–6790\,\AA\, (red). The combination of high multiplexing, broad spectral coverage, and dual-resolution modes will enable transformative science on WD+MS binaries, providing precise white dwarf ages to calibrate age–metallicity and age–velocity dispersion relations in the Milky Way, and to establish age–rotation–activity relations for low-mass main-sequence stars, thereby linking stellar evolution with Galactic structure and dynamics.

\vspace{0.5cm}

\emph{This version contains an abstract, an Acknowledgments section and a longer bibliography than the one submitted to ESO.}

\section{Acknowledgments}

This work was partially supported by the MINECO grant PID2023-148661NB-I00 and by the AGAUR/Generalitat de Catalunya grant SGR-386/2021, and by the MINECO grant PID2020-117252GB-I00 and the PhD grant PRE2021-100503 funded by MICIU/AEI/10.13039/501100011033 and ESF+. RR acknowledges support from Grant RYC2021-030837-I, funded by MCIN/AEI/ 10.13039/501100011033 and by “European Union NextGeneration EU/PRTR”. MC acknowledges grant RYC2021-032721-I, funded by MCIN/AEI/10.13039/501100011033 and by the European Union NextGenerationEU/PRTR. EMGZ acknowledges financial support from Banco de Santander, under a Becas Santander Investigación/Ajuts de Formació de Professorat Universitari (2022\_FPU-UPC\_16). AFiB acknowledges financial support from Ajut predoctoral cofinançat per la Unió Europea Joan Oró    2024\,FI-1\,00010 grant.

\end{document}